\documentclass[pra,twocolumn,showpacs,preprintnumbers,amsmath,amssymb]{revtex4-1}
\usepackage{graphicx}% Include figure files
\usepackage{dcolumn}% Align table columns on decimal point
\usepackage{bm}% bold math
\usepackage{amsmath}
\DeclareGraphicsExtensions{.pdf,.eps,.png,.jpg,.mps} 

\begin{document}

\preprint{}

\title{Spin and charge caloritronics in bilayer graphene flakes with magnetic contacts}

\author{Leonor Chico}
\affiliation{ Instituto de Ciencia de Materiales de Madrid, Consejo Superior de Investigaciones Cient\'\i ficas, C/ Sor Juana In\'es de la Cruz 3, 28049 Madrid, Spain}

\author{P.A. Orellana}
\affiliation{Departamento de F\'{i}sica, Universidad T\'{e}cnica
Federico Santa Mar\'{i}a, Casilla 110-V, Valpara\'{i}so, Chile}

\author{L. Rosales}
\affiliation{Departamento de F\'{i}sica, Universidad T\'{e}cnica
Federico Santa Mar\'{i}a, Casilla 110-V, Valpara\'{i}so, Chile}

\author{M. Pacheco}
\email{monica.pacheco@usm.cl}
\affiliation{Departamento de F\'{i}sica, Universidad T\'{e}cnica
Federico Santa Mar\'{i}a, Casilla 110-V, Valpara\'{i}so, Chile}

\date{\today}

\begin{abstract}

We investigate the coupling of spin and thermal currents as a means to rise the thermoelectric efficiency of nanoscale graphene devices.  We consider  nanostructures composed of overlapping graphene nanoribbons with ferromagnetic contacts in different magnetic configurations.  
 Our results show that the charge Seebeck effect is greatly enhanced when the magnetic leads are in an antiparallel configuration, due to the enlargement of the transport gap. However, for the optimization of the charge figure of merit ZT it is better to choose a parallel alignment of the magnetization in the leads, because the electron-hole symmetry is broken in this magnetic configuration. We also obtain the spin-dependent Seebeck coefficient and spin figure of merit. In fact, the spin ZT  
can double its value with respect to the charge ZT for a wide temperature range, above 300 K. 
 These findings suggest the potential value of graphene nanosystems as energy harvesting devices employing spin currents. 
\end{abstract}

\maketitle

\section{Introduction} 

Thermoelectric materials allow for the conversion of heat into electricity and vice versa. Besides the fundamental interest in thermoelectric properties, that can be analyzed from the general viewpoint of transport phenomena, their potential applications for energy recovery or harvesting in processes where a large amount of heat is dissipated are one of the main motivations for their study \cite{Tritt2011}. 

The efficiency  of  thermoelectric devices is  related to their figure of merit, defined as $ZT=S^2 G T / \kappa$ where $S$ is the Seebeck coefficient or thermopower,  $ G $ is the electronic conductance, $ \kappa$ the thermal conductance and $T$ is the temperature. In principle, the Seebeck coefficient is large for insulators, whereas  a high electronic conductance is obtained in conductors, so usually these two magnitudes are related in a conflicting way. Additionally, a low thermal conductance $\kappa$ would also enhance $ZT$, which includes an electronic and a phononic contribution.  A possible way to
tune these properties in an independent fashion is to lower the dimension and go to the nanoscale \cite{Kim2009,Ouyang2009,Dubi2011}. 

In the last years the thermoelectric properties of nanostructured materials have been investigated with the aim to enhance  thermoelectric efficiencies. Quantum size effects permit the variation of the electronic properties \cite{Finch2009,GomezSilva2012,Trocha2012,GarciaSuarez2013,Cortes2017} and the reduction of the phonon thermal conductance \cite{Zhou2014}. Nanopatterning, either with antidots, defects or edge modification, allows for further suppression of phonons \cite{Karamitaheri2011,Chang2012}. So despite the fact that pristine graphene is not a good material for thermoelectrics, being gapless and very good thermal conductor, graphene nanostructures have been proposed as potentially interesting for thermoelectric applications \cite{Dollfus2015}, with excellent values of the thermopower coefficient and figure of merit \cite{Mazzamuto2011,Haskins2011,Sevincli2013,Rosales2013,Li2014,Cortes2014,SaizBretin2015,Hossain2016}.  

One way to increase the thermopower is to enhance the electron-hole asymmetry at the Fermi energy \cite{Walter2011,Hossain2016}. 
Strong asymmetries can be found in ferromagnetic materials, so 
the coupling of the spin and thermal currents have been proposed as a means to further improve the thermoelectric properties of nanoscale devices \cite{Slachter2010,Walter2011,Slachter2011,Bauer2012}. In such systems, 
charge, spin and heat transport are related, and may provide opportunities to optimize their performance. This active field has been called spin caloritronics \cite{Bauer2010}; although its bases were settled in the pioneering work by Johnson and Silsbee \cite{Johnson1987}, its development in low-dimensional systems has 
boosted the interest in these phenomena. Spin-dependent thermoelectric properties of magnetic graphene have been previously studied \cite{Rameshti2015}, but without taking advantage of the reduced dimensionality and neglecting the phonon contribution to the thermal conductance.

In this work, we explore the spin-dependent thermoelectric properties of quasi-one-dimensional graphene structures composed of overlapping nanoribbons. 
Such systems were studied before due to their remarkable transport properties; if the ribbons are metallic, a series of resonances and antiresonances appear 
related to the length of the bilayer region \cite{Jhon2010,Jhon2011}. If the ribbons are semiconducting, they present important advantages due to the reduction of the thermal conductance in roughly an order of magnitude, yielding an enhanced figure of merit  \cite{Nguyen2014}. This is due to the fact that 
only van der Waals forces couple the bilayer central part, so the strong graphene $\sigma$-bond lattice is interrupted in the overlapping region, hindering phonon transport. Here we explore the advantages of considering spin-dependent transport in the thermoelectric properties; specifically, the role of magnetic contacts in the 
spin-thermoelectric response of these structures. 

Our main results are the following:\\
(1)  The charge Seebeck coefficient and figure of merit are strongly dependent of the magnetic configuration of the leads. We find that the charge Seebeck coefficient is greatly enhanced when the magnetic leads are in an antiparallel configuration, due to the enlargement of the transport gap.\\
(2)  Conversely, the charge figure of merit improves if the magnetizations of the leads are parallel, due to the electron-hole asymmetry.\\
(3) Large values for the spin-dependent Seebeck coefficient can be achieved for a wide range of energies, up to twice the charge Seebeck coefficient. Due to the changes on the spin-dependent Seebeck coefficient,  the spin figure of merit may have more than one maximum with an opposite temperature dependence. 

In Section II of the paper we describe the theory and model we employed to calculate the charge and spin-dependent thermoelectric coefficients of the bilayer graphene structure.  In Section III  we present results for the conductances,  Seebeck coefficients and figure of merits  as functions of the chemical potential. 
Finally, Section IV summarizes our conclusions.

\section{System and Model} \label{teorico}

\subsection{Geometry and configuration of the system}

\begin{figure}[h!]
\centering
\includegraphics[width=90mm,clip]{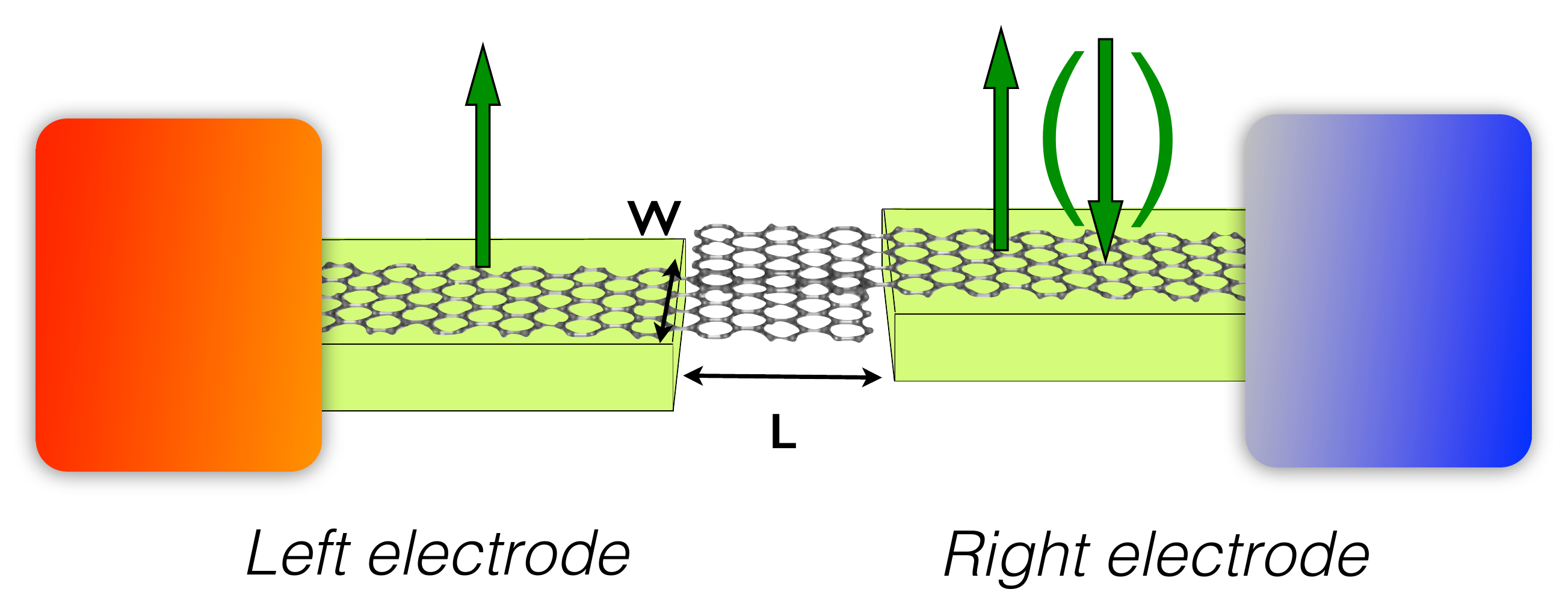}
 \caption{Depiction of the graphene flake of length $L$ and width $W$  made by two overlapping nanoribbons. The left and right leads, i.e., the monolayer ribbons, are deposited on top of a ferromagnetic material. The magnetic moment of the right lead may be parallel or antiparallel to that of the left lead. A difference of temperature between the electrodes is applied. }
  \label{figflake}
\end{figure}

As commented in the Introduction, we consider two
overlapping armchair nanoribbons with AA stacking. 
 The geometry is shown in Fig.~\ref{figflake}. This system can be alternatively viewed as a finite-size bilayer flake connected to two monolayer ribbons acting as electrodes, which can be ferromagnetic due to proximity effects \cite{Lee2010,WangTang2015}. We take into account two configurations for the leads, with the magnetic moments of the left and right lead being parallel (P) or antiparallel (AP), as depicted in the figure. There is a temperature difference between the monolayer electrodes, giving rise to thermoelectric effects.

\subsection{Model} 

We employ a  $\pi$-orbital tight-binding model, which gives an excellent description of the electronic properties of graphene systems around the Fermi level. 
 The in-plane nearest-neighbor interaction is given by a single hopping parameter $\gamma$, which we take as $-3$ eV. The interlayer coupling in the flakes is considered by one hopping parameter between atoms directly placed on top of each other, being  $\gamma^{\prime}=0.1 \gamma$. The ferromagnetic electrodes are modeled by the inclusion of a spin-dependent onsite term $\Delta=\pm 0.1$ eV, which produces a 0.2 eV splitting between majority and minority bands in the leads.
 
As it is customary, armchair (AC) ribbons are labeled with the number of dimers $n$ across its width, i.e., AC$n$. Thus, an AC12 ribbon has a width of 13.53 
${\rm \AA}$. The length of the flake is given in terms of the number $m$ of translational unit cells. The length of one unit cell for this type of ribbon is equal to 3$a_c$, being $a_c=1.42$ ${\rm \AA}$ the carbon-carbon bond length. In this way we can identify the systems as AC$n$L$m$, i.e., giving the width of the ribbon and the length of the flake. 
 
\begin{figure}[ht]
\centering
\includegraphics[width=\columnwidth,clip]{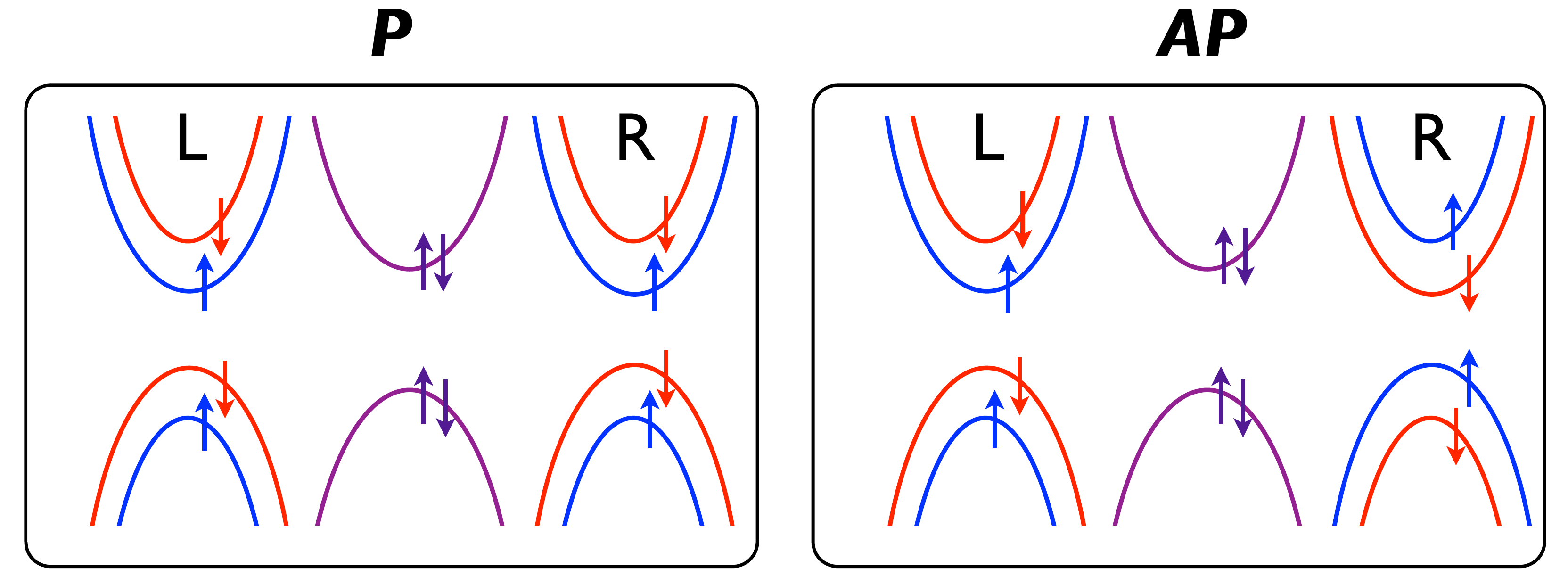}
\caption{ Schematic band structures of the left lead, central part (flake) and right lead in the P (left panel) and AP (right panel) magnetic configurations.} 
\label{figbands}
\end{figure}

The electronic transmission $\tau$ as a function of the energy carrier $E$ through the system can be numerically obtained by standard Green function techniques, following a real-space renormalization scheme \cite{Chico1996b}. From the transmission all the thermoelectric magnitudes of interest can be computed. 
We neglect spin-scattering effects, meaning that each spin channel contributes independently to the conductance and spin-flip processes are neglected. 
As the right (R) and left (L) leads can be in two different ferromagnetic states, we have two different situations with respect to the spin-dependent transport through the flake, depicted in Fig.~\ref{figbands}. In the parallel configuration (P), the band structure of the left and right lead are exactly equal, whereas in the AP case they are not.

\subsection{Spin-dependent thermoelectric magnitudes}

  The spin-dependent electric current can be obtained in terms of the transmission function within the Landauer-B\"uttiker formalism
\begin{equation}
I_{\sigma} = \frac{e}{ h} \int_{-\infty} ^\infty   \tau_{\sigma}(E)(f_L(E)-f_R(E))dE\,\,\,,
\end{equation}
where $f_{R,L}$ are the Fermi distributions of the right and left electrodes and $\sigma=\uparrow (\downarrow)$ denotes the spin projection direction.

The spin-resolved electronic conductance is given by
$G_{\sigma}=\Delta I_{\sigma}/\Delta V$, in the limit of  $\Delta V\rightarrow 0$.

Analogously, the spin-resolved thermal current is given by  
\begin{equation}
I_{Q,\sigma} =  \int_{-\infty} ^\infty  (E-\mu+\Delta V) \tau_{\sigma}(E)(f_L(E)-f_R(E))dE\,\,\,,
\end{equation}

The Seebeck coefficient or thermopower $S$ is defined as the voltage drop induced by a temperature gradient at vanishing current, 
$S=-\Delta V/\Delta T|_{I_{\sigma}=0}$, 
in the limit of  $\Delta T\rightarrow 0$.  The electronic thermal conductance defined as 
$\kappa_{e,\sigma}=\Delta I_{Q,\sigma}/\Delta V$ for $\Delta V\rightarrow 0$
 is calculated in the linear response regime i.e., $|\Delta T|<<T$ and $|e\Delta V|<<\mu$, with $\mu$ being the equilibrium chemical potential, and $T$ the temperature.

The above-defined thermoelectric coefficients  can be written in compact form in terms of the following thermal integrals:

\begin{equation}
{\cal L}_{n,\sigma}(\mu,T)= \int_{-\infty} ^\infty x^n\frac{e^x}{\left(e^x+1\right)^2}   \tau_{\sigma}(x,\mu,T) \, dx . 
\label{Ln}
\end{equation}

Thus the electronic conductance is just 
\begin{equation}
G_\sigma=\frac{e^2}{h}{\cal L}_{0,\sigma},
\end{equation}
where $e$ is the elementary charge and $h$ the Planck's constant; 
the Seebeck coefficient is given by 

 \begin{equation}
 S_\sigma= -\frac{k_B}{e}\frac{{\cal L}_{1,\sigma}}{{\cal L}_{0,\sigma}},
 \label{Ssig}
 \end{equation}
with $k_B$ being the Boltzmann's constant; and finally, the electronic thermal conductance is

 \begin{equation}
 \kappa_{e,\sigma}= \frac{k_B^2T}{h}\left[ {\cal L}_{2,\sigma} - \frac{{\cal L}_{1,\sigma}^2}{{\cal L}_{0,\sigma}} \right].
 \end{equation}

 The charge and spin conductance are defined as $G_{c}=G_{\uparrow}+G_{\downarrow}$ and $G_{s}=G_{\uparrow}-G_{\downarrow}$,
 and the charge Seebeck and spin-dependent Seebeck coefficients  as $S_{c}=(S_{\uparrow}+S_{\downarrow})/2$ and $S_{s}=S_{\uparrow}-S_{\downarrow}$, respectively \cite{Slachter2010}.

With these definitions, we can obtain the corresponding charge and spin  figures of merit, 
\begin{equation}
Z_{c,s}T=\frac{G_{c,s} S_{c,s}^2 T}{\kappa}
\end{equation}
\noindent
 where $\kappa=\kappa_{e\uparrow}+\kappa_{e\downarrow}+\kappa_{ph}$, with $\kappa_{e\sigma}$ and $\kappa_{ph}$ are  the electronic and phonon contribution to the charge thermal conductance, respectively. 
 
  In what follows we show results for charge-dependent magnitudes $G_{c}$, $S_{c}$ and $Z_{c}T$ and spin-dependent magnitudes   $G_{s}$, $S_{s}$.   Note that in the AP configuration  $G_s=0$, $S_s=0$ and  $Z_sT=0$ due to the symmetry of the spin-split bands in the leads, as can be inferred from Fig. \ref{figbands}. 

\section{Results and discussion} 

As the main ingredients to compute spin-thermoelectric magnitudes are the spin-resolved transmissions, we first show these values for the systems studied in this work.
We choose the ribbons to be semiconducting, because they yield better values of the thermoelectric properties than the metallic counterparts  \cite{Nguyen2014}, for which a fine tuning
of their geometry is needed to increase their thermoelectric response \cite{Cortes2017}. We have verified that certainly  is more interesting to focus in semiconducting flakes. 

\begin{figure}[h!]
\centering
\includegraphics[width=\columnwidth,clip]{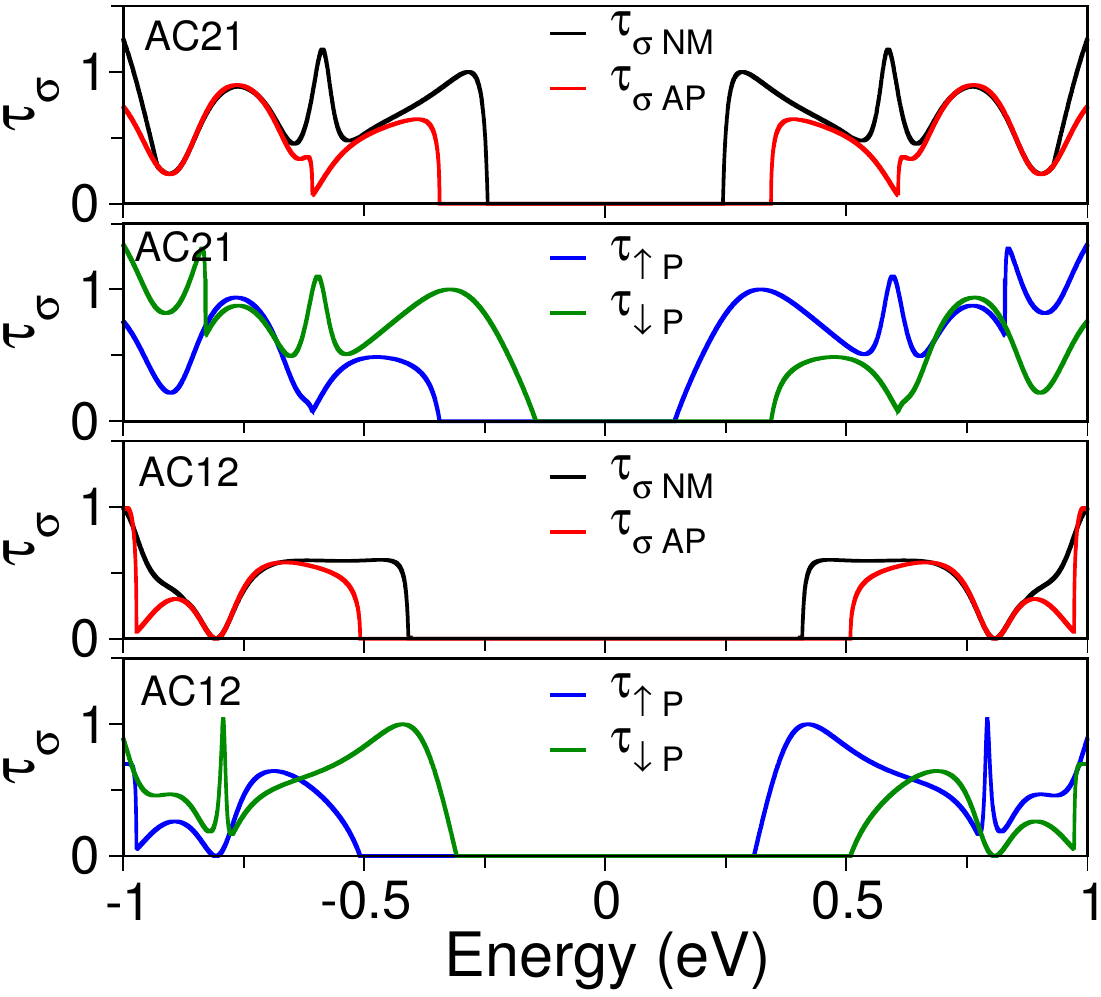} 
\caption{Spin-resolved electronic transmissions for AC12L12 and AC21L12 graphene flakes  in the non-magnetic (black), P (blue for majority spin, green for minority spin) and AP (red) configurations.} 
\label{figtau}
\end{figure}

Fig. \ref{figtau} shows the spin-resolved transmissions for  the AC21L12 and AC12L12 flakes in the AP and P configurations; for the sake of comparison, we include also the transmission in the case of non-magnetic (NM) leads. As mentioned before, the transmission for the AP configuration is the same for both spin directions (see Fig.~\ref{figbands}); obviously, this also happens trivially for the NM case. The asymmetry in the transmission for the P configuration is noteworthy; this can be a way to increase the thermoelectric response.
Nonetheless, there is a clear relation between  $\tau_\uparrow$  and $\tau_\downarrow$ in the P configuration due to the electron-hole symmetry present in graphene systems, namely,  $\tau_\uparrow^P(E) = \tau_\downarrow^P(-E)$.  
Notice that the conductance gap increases in the AP configuration; however, for the P case, the gap for each spin channel is unchanged in comparison to the NM case, although the {\em total} transmission 
has a smaller conductance gap, as it can be clearly seen in Fig. \ref{figtau}. 

\begin{figure}[!ht]
\centering
\includegraphics[width=\columnwidth,clip]{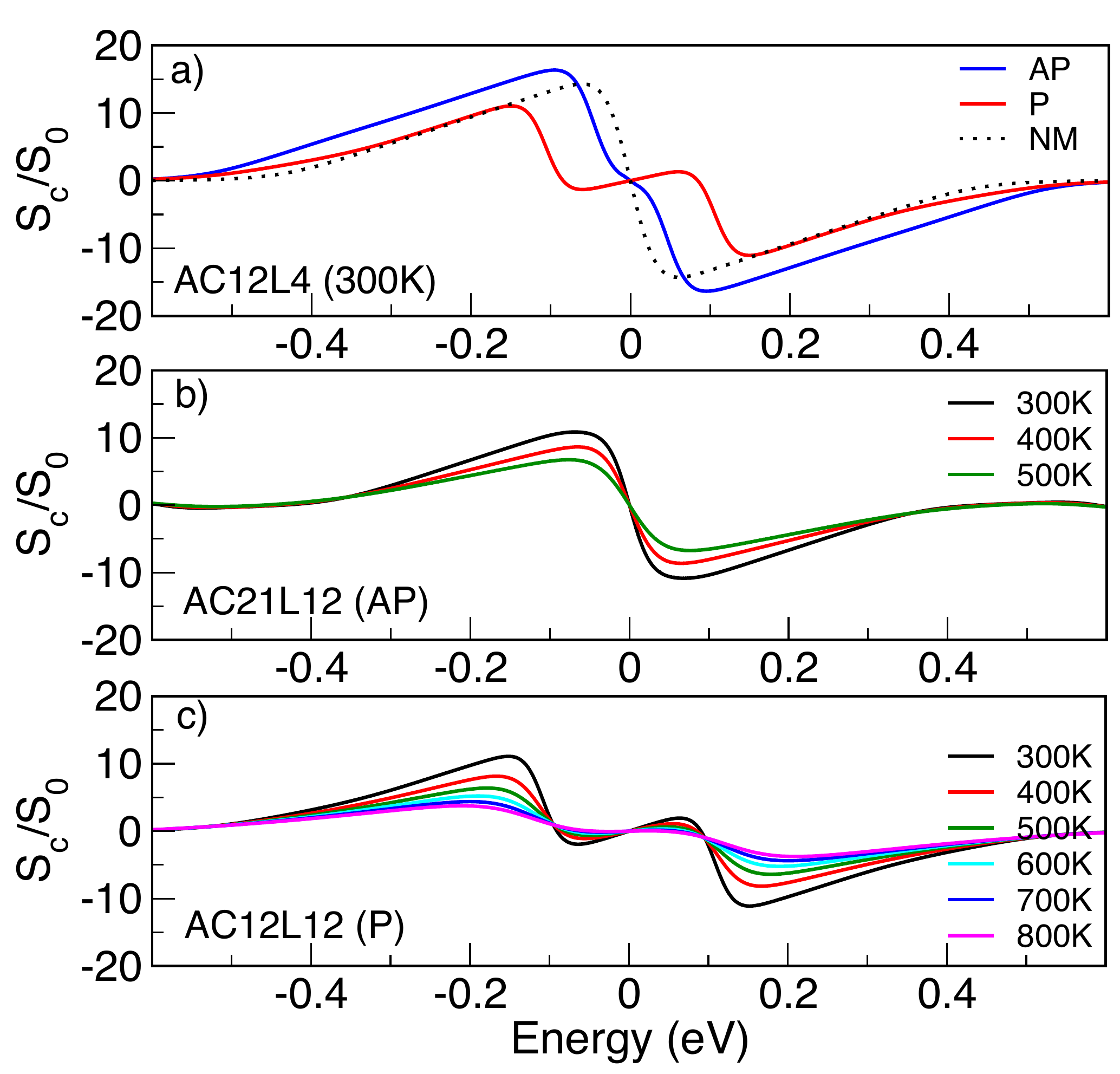}\caption{ a) Charge Seebeck coefficients of the AC12L4 system at $T=300$ K, for the P, AP and NM configurations.
b) Temperature dependence of the charge Seebeck coefficient for the AC21L12 case in the AP configuration and c) temperature dependence of the charge Seebeck coefficient for the AC12L12 case in the P configuration.} 
\label{figseebch}
\end{figure}

In Fig. \ref{figseebch} (a) we present the charge Seebeck coefficients (in units of $S_0=k_B/e$) for the three magnetic configurations, NM, P and AP, for the AC12L4 case, at $T=300$ K. 
The optimal charge thermopower is obtained for the AP configuration, being better than the NM case. This is due to the increase of the gap for the AP alignment. The smallest $S_c$ is obtained for the P configuration; this is due to the reduction of the conductance gap. As $S_c$ is the sum of the spin-up and down contributions, 
the P charge thermopower has more structure due to the addition of the two displaced $S_\uparrow$ and $S_\downarrow$ spin parts, shifted in 0.1 eV with respect to the NM case. 
The reduction of $S_c$ has also been found in quantum dots with ferromagnetic leads with parallel magnetization \cite{Trocha2012}.
We have checked that  $S_c$ is almost unchanged with respect to the flake length for all the magnetic configurations. This can be understood because the 
Seebeck coefficient is mainly dependent on the transmission gap, for which the flake characteristics are not relevant. Indeed, our $S_c$ for the NM configuration is similar to that reported in Ref. \cite{Ouyang2009} for the corresponding perfect ribbons.
 
In order to inspect the role of the nanoribbon width, we show the temperature dependence for the AC21L12 flake in the AP magnetic configuration in the  panel of Fig.   \ref{figseebch} (b). This is a wider ribbon, so its  gap is smaller; for this reason the Seebeck coefficients are reduced with respect to the AC12 flakes. We observe a decrease of the thermopower with increasing temperature for both magnetic configurations; Fig.\ref{figseebch} (c) shows a wide range of temperatures for the AC12L12 system in the P magnetic configuration. This is consistent with previous calculations in pristine semiconducting nanoribbons \cite{Ouyang2009}.

For the sake of obtaining a reliable quantity for the figure of merit, it is necessary to include the phonon thermal conductance. We draw these values from the literature. In Ref. \cite{Nguyen2014} the same geometry (AC12) for different flake lengths  is investigated, albeit without including the spin dependence. 
In their Fig. 3 (a) they present the phonon conductance as a function of temperature. For temperatures above 300K, the phonon conductance is almost constant, so we take their values at 600 K for all the temperatures studied, namely,  $\kappa_{ph}=0.28$ nW/K for L=12,  $ \kappa_{ph}=0.22$ nW/K for L=8, and $ \kappa_{ph}=0.15$ nW/K for L=4. In all cases these are reasonable upper bounds for the phonon thermal conductance.  For the wider flake, AC21L12, we extrapolate the value from the calculated phonon thermal conductance presented in Ref. \cite{Huang2010}; we take $\kappa_{ph}=0.38$ nW/K.
 
\begin{figure}[h!]
\centering
\includegraphics[width=\columnwidth,clip]{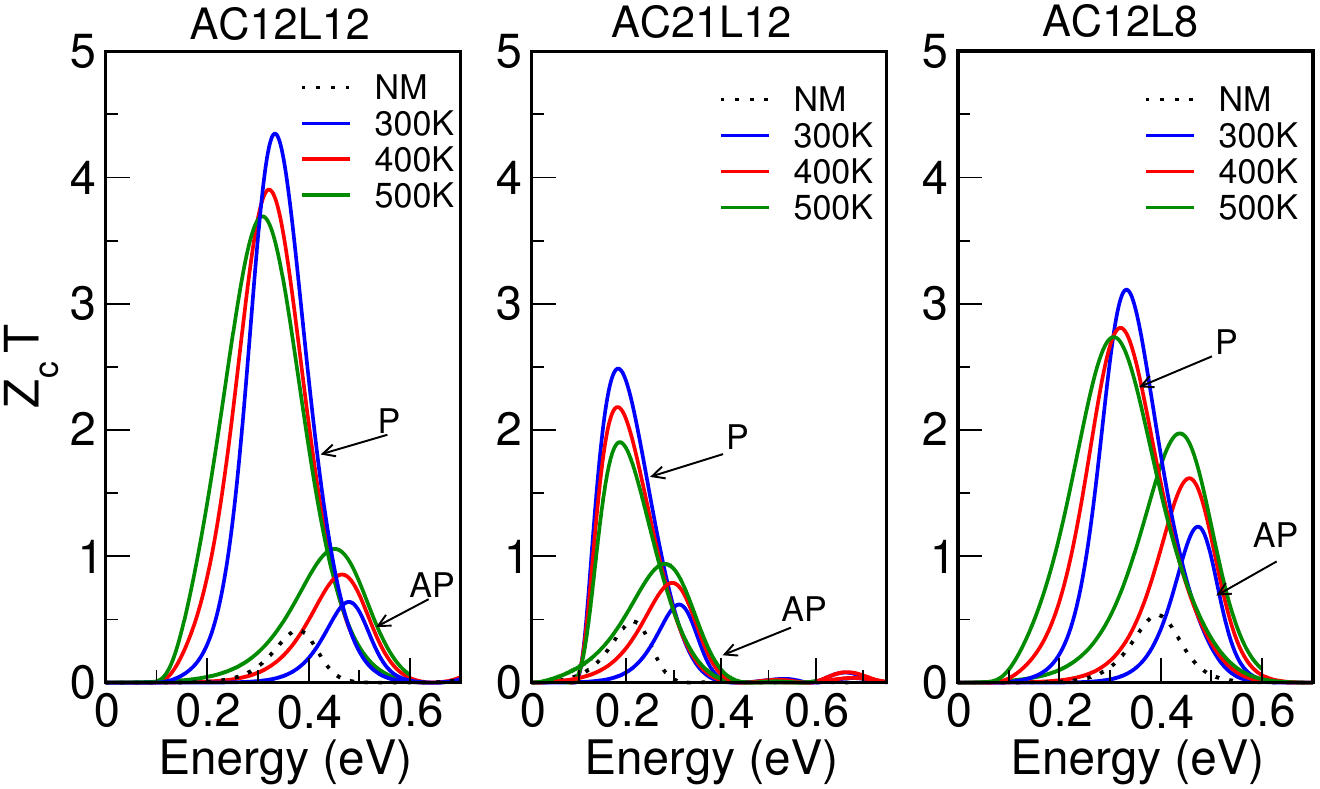}
\caption{Charge ZT for AC12L12, AC21L12 and AC12L8 flakes in the P and  AP  magnetic configurations for different temperatures. The NM case is included in black dotted line for 300K.
 }
 \label{figztch}
\end{figure}

In Fig. \ref{figztch} we present our results for the charge figure of merit $Z_cT$ corresponding to the flakes AC12L12, AC12L8 and AC21L12 in the P and AP magnetic configurations and three different temperatures, 300 K, 400 K and 500 K. For comparison, we have also plotted the  results for nonmagnetic  (NM) leads in black dotted lines for 300 K.
We observe that for all the systems the maximum value of $Z_cT$ ($Z_cT_{\rm max}$) is greatly enhanced if the ferromagnetic leads are in the parallel magnetic configuration, 
while in the antiparallel case  $Z_cT_{\rm max}$ takes values slightly greater than those for NM leads. This behavior is a consequence of the decrease in the transmission gap in the P configuration, as it can be observed in Fig. \ref{figtau}.  The increase in the electronic conductance permits to reach maximal intensities for the figure of merit, despite the fact that  this is not the optimal magnetic arrangement  to obtain high Seebeck coefficients. 
For the range of temperatures studied,  $Z_cT_{\rm max}$ in the  AP configuration increases with temperature. This is an interesting behavior which we also observe for the NM case.  It is a consequence of the increase in the electronic conductance with $T$ and the
% almost constant value 
independence of the phonon thermal conductance
for $T\geq 300$ K  in these structures (see Ref. \cite{Nguyen2014}). In  the P configuration, $Z_cT$ has a dual behavior with T as a function of the chemical potential.  
 Depending on the energy range, 
  the figure of merit decreases with increasing temperature, as expected in a normal semiconductor device,
  whereas in other energy intervals the behavior is the opposite. This is most clearly seen in the AC12 cases (the left and right panels of Fig. \ref{figztch}).   
  Such changes in the temperature dependence are very sensitive to the gap value and spin splitting energy $\Delta$. 
  
\begin{figure}
\centering
\includegraphics[width=\columnwidth,clip]{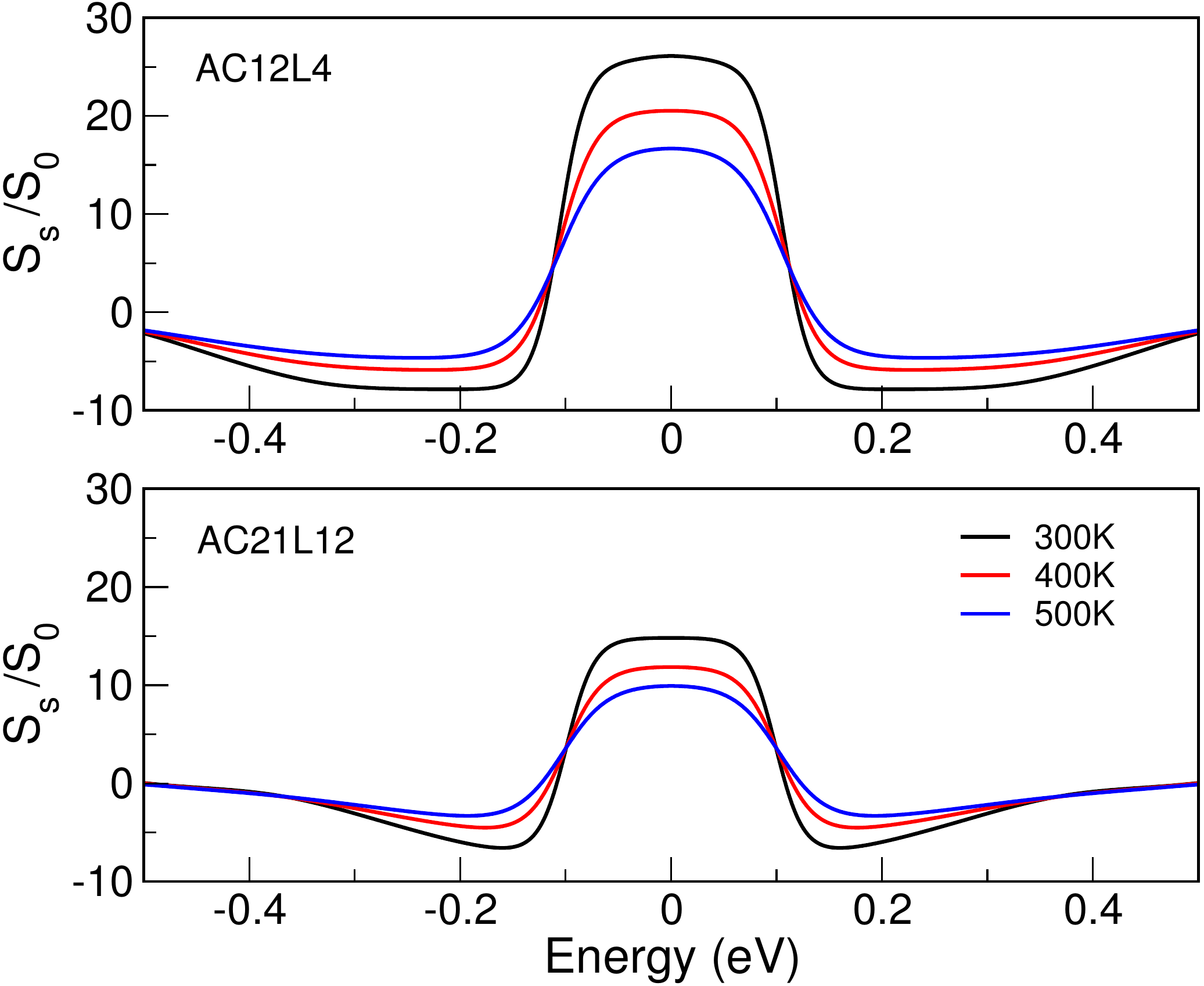}
\caption{Spin-dependent Seebeck coefficient 
 for AC12L12 and AC21L12 for different temperatures.}
\label{fig6}
\end{figure}

In  Fig. \ref{fig6} we present the spin-dependent Seebeck coefficients for  the AC12L4 (upper panel) and the ACL21L12 (bottom panel) systems  for three different temperatures. As the spin-dependent thermopower is the subtraction of the Seebeck coefficients of the two spin channels, for a nonzero $\Delta$ there will always be a finite contribution due to the breaking of electron-hole symmetry. For larger spin splitting $\Delta>0.3$ eV, the $S_s$ will have a distinct shape, being possible to discern the two contributions of the two spin channels separately. However, for $\Delta=0.1$ eV, chosen in this work, we have a plateau around 0 eV due to the overlap of the Seebeck coefficients of the two spin channels. The maximum  $S_s$ is considerably high, slightly larger than that obtained for the charge Seebeck coefficient.  

If electron-symmetry holds, the charge Seebeck coefficient is an odd function of the energy, as it is evident from Fig. \ref{figseebch} and readily inferred from Eqs. \ref{Ln} and \ref{Ssig}. Since in our system in the P configuration the spin splittings at the leads have the same value, the spin-dependent Seebeck coefficient is also symmetric, although now it is an even function of the energy, having at least two zeros close to  $\pm \Delta$.

 \begin{figure}
\centering
\includegraphics[width=\columnwidth,clip]{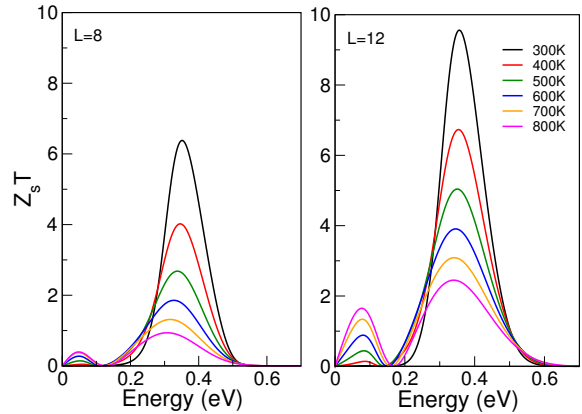}
\caption{  Spin-dependent figure of merit for AC12L12 and AC12L8 for several temperatures.}
 \label{fig7}
\end{figure}

In Fig. \ref{fig7} we present our results for the spin-dependent figure of merit $Z_sT$ for the flakes AC12L12, AC12L8 in the temperature interval from 300 K to 800 K.
There are two peaks related to the two intervals of positive and negative values of $S_s$, which are shifted with respect to the non-magnetic case. Although the maximum absolute value of the spin-dependent Seebeck coefficient is attained in the central plateau, the role of the electronic conductance is crucial to set the maximum 
value of  $Z_sT$, $Z_sT_{\rm max}$. Notice that  $Z_sT_{\rm max}$ takes place near the band edges, boosting up to twice the value of the charge figure of merit (see Fig. \ref{figztch}).
The two peaks have quite dissimilar value, due to the electronic conductance, and also exhibit an opposite temperature dependence, which changes at the minimum imposed by the zero of the spin-dependent Seebeck coefficient. 

These findings indicate that it is advantageous to exploit spin currents with the aim of maximizing the figure of merit, especially for nanostructured materials. Notice that similar effects to those achieved in the AP configuration can be obtained by applying a gate voltage at the leads: if the chemical potential at the electrodes if varied, one can mimic the role of the ferromagnetic leads analyzed in this work, therefore tuning by electrical means the thermoelectric behavior of the system. 

\section{Conclusions}

We have shown that the coupling of spin with charge and thermal currents can be harnessed to rise the thermoelectric efficiency of nanoscale graphene devices.  The use of magnetic leads in  nanostructures  composed of overlapping graphene nanoribbons allows for the optimization of the {\em charge} thermoelectric properties, as well as the spin-dependent magnitudes. 

We have demonstrated  that the charge Seebeck effect is greatly enhanced when the magnetic leads are in an antiparallel configuration, due to the enlargement of the transport gap. However, for the optimization of the charge figure of merit, it is better to choose a parallel alignment of the magnetization in the leads, due to 
the electron-hole asymmetry present in this configuration. We also obtained the spin-dependent Seebeck coefficient and figure of merit. In fact, the latter shows much better values
 than their charge counterparts, up to 10 at room temperature. 
 These findings suggest the potential value of graphene nanosystems as energy harvesting devices employing spin currents and magnetic electrodes.

\acknowledgments
We acknowledge helpful discussions with P. Vargas. 
This work has been partially supported by the Spanish MINECO under 
grant FIS2015-64654-P, Chilean FONDECYT grants 1140571 (P.O.), 1140388 (L. R.), 1151316 (M.P.).
L. C. gratefully thanks the hospitality of the Universidad T\'ecnica Federico Santa Mar\'{\i}a (Chile). 

%\bibliography{ref_calor}


\end{document}